\documentclass{article}
\usepackage{spconf,amsmath,graphicx, amsfonts}

\usepackage{enumitem}
\setlist{nosep, leftmargin=14pt}

\usepackage{mwe} 

\title{3D Coronary Vessel Reconstruction from Bi-Plane Angiography using Graph Convolutional Networks}

\name{%
\begin{tabular}{@{}c@{}}
Kit Mills Bransby\textsuperscript{1} \qquad 
Vincenzo Tufaro\textsuperscript{1,2} \qquad 
Murat \c Cap\textsuperscript{1,2} \\ 
Greg Slabaugh\textsuperscript{1} \qquad 
Christos Bourantas\textsuperscript{1,2} \qquad 
Qianni Zhang\textsuperscript{1}
\end{tabular}}

\address{\textsuperscript{1} Queen Mary University of London, United Kingdom \\
\textsuperscript{2} Department of Cardiology, Barts Health NHS Trust, London, United Kingdom 
}

\begin{document}
\ninept
\maketitle
\begin{abstract}
X-ray coronary angiography (XCA) is used to assess coronary artery disease and provides valuable information on lesion morphology and severity. However, XCA images are 2D and therefore limit visualisation of the vessel. 3D reconstruction of coronary vessels is possible using multiple views, however lumen border detection in current software is performed manually resulting in limited reproducibility and slow processing time. In this study we propose 3DAngioNet, a novel deep learning (DL) system that enables rapid 3D vessel mesh reconstruction using 2D XCA images from two views. Our approach learns a coarse mesh template using an EfficientB3-UNet segmentation network and projection geometries, and deforms it using a graph convolutional network. 3DAngioNet outperforms similar automated reconstruction methods, offers improved efficiency, and enables modelling of bifurcated vessels. The approach was validated using state-of-the-art software verified by skilled cardiologists. 
\end{abstract}
\begin{keywords}
\ninept deep learning, 3d reconstruction, angiography
\end{keywords}
\section{Introduction}
\label{sec:intro}

\ninept XCA is a standard procedure in the assessment of coronary artery disease, where an injection of radiopaque contrast medium into vessels enables visualisation of lumen morphology \cite{garrone2009quantitative}. Regions of narrowing (stenosis) caused by a build up of atherosclerotic plaque that restrict blood flow to the heart can be identified, aiding treatment such as stent placement. Quantitative coronary angiography (QCA) was introduced to provide precise quantification of plaque lesions and disease progression. However, QCA only uses a 2D representation of the lumen, and therefore limits visualisations in cases of vessel overlap and foreshortening where the full morphology of the vessel is either obscured or distorted \cite{garrone2009quantitative}. In order to overcome these problems, 3D-QCA methods have been developed to reconstruct coronary artery segments in three dimensions using two or more angiographic views. 3D-QCA models are of high clinical relevance due to their application in computational fluid dynamics where the flow of blood through coronary vessels is simulated. These approaches allow for visualisation of vessel geometry and stenosis in 3D, however they are typically semi-autonomous, time-consuming, and require manual correction of vessel segmentation or multi-step input from clinicians. Bifurcation points where the vessel splits into multiple branches affect the flow and velocity of blood, however reconstructing main and side branches in a single 3D-QCA model has been a particular challenge in past investigations. 
\begin{figure*}[ht]
\centering
\includegraphics[width=1\textwidth]{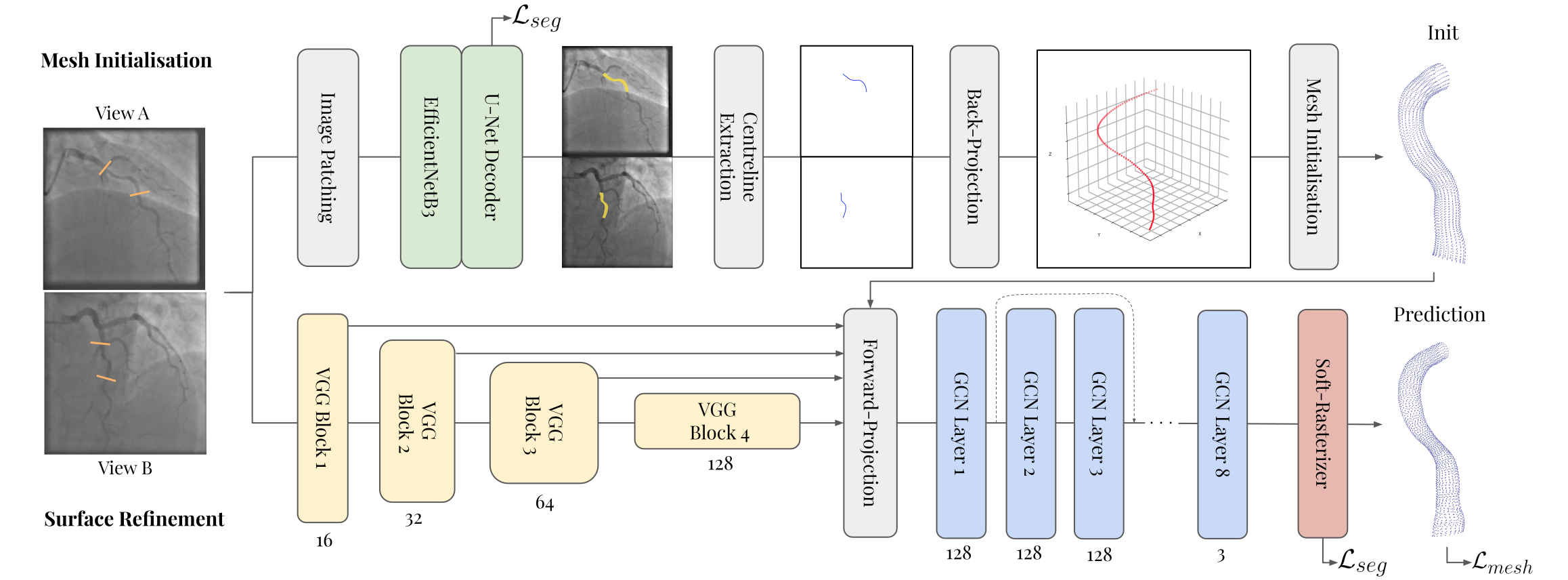} \caption{3DAngioNet framework.} \label{pipeline}
\end{figure*}
\\
\indent Since the inception of the ShapeNet dataset \cite{shapenet2015}, mesh deformation DL algorithms such as 3DR2N2 \cite{choy20163d}, Pixel2Mesh \cite{wang2018pixel2mesh}, Pixel2Mesh++ \cite{wen2019pixel2mesh++} have demonstrated state-of-the-art results for 3D reconstruction from 2D images. This is especially applicable in medical imaging where data is often limited to one or few views and has led to 3D reconstruction of the heart in HeartFFDNet \cite{kong2021deep}, liver in Xray2Shape \cite{tong2020x} and lungs in DeepOrganNet \cite{wang2019deeporgannet}. Such methods offer advantages over traditional approaches as they are able to reconcile surface morphology in information poor areas, have higher performance and are computationally efficient. Despite this, no attempts have been made to reconstruct coronary vessels using these advanced DL techniques.
\\
\indent \textbf{Our Contribution.} Based on the challenges discussed, we propose and validate a novel deep learning methodology for 3D reconstruction of coronary artery segments called 3DAngioNet. To the best of our knowledge this is the first deep learning paper to automate coronary segment reconstruction with bifurcation points and without the need for clinical correction.

\section{method}
\label{sec:method}

\subsection{System Overview}
\label{ssec:overview}
3DAngioNet uses a coarse-to-fine approach based on three modules. Firstly, a Mesh Initialisation (MI) module creates a coarse mesh using an EfficientNetB3-UNet segmentation algorithm and back-projection from 2D to 3D using stereo geometry. Secondly, a Surface Refinement (SR) module adds fine-grained detail by deforming the initial mesh using image features sampled from an encoder using a graph convolutional network (GCN). Third, in cases of bifurcation an additional step is carried out where main and side branches are stitched together using a simple Boolean operation. The model takes a pair of angiographic images as input, with corresponding acquisition geometries and segment of interest (SOI) start and end points provided by an expert clinician, and outputs a 3D mesh surface. The full framework is presented in Fig. \ref{pipeline}.

\subsection{Mesh Initialisation Module}
\label{ssec:meshinitmodule}
\textbf{2D Vessel Segmentation.} As segmentation of a specific SOI is required, there are a number of challenges to overcome when using deep learning segmentation algorithms. Firstly, as the SOI is similar to other vessels in shape, texture and pixel values, it is likely a DL segmentation algorithm will try to segment the full coronary tree resulting in a high number of false positives. This was confirmed experimentally and overcome by cropping the X-ray around the SOI. Despite this, overlapping vessels in the cropped patch may still restrict learning as the choice of vessel can be ambiguous. This was mitigated by rotating images so the start and end points of the SOI are fitted to the x-axis. This strategy ensures the algorithm learns to segment a vessel only if it is orientated in a horizontal position, vastly reducing false positives. A U-Net model \cite{ronneberger2015u} with an EfficientNetB3 \cite{tan2019efficientnet} encoder backbone initialised with ImageNet weights is used as the segmentation architecture. The training of the network is supervised by a Binary Cross Entropy Dice loss ($\mathcal{L}_{seg}$) between the predicted probability map and a binary ground truth map. 
\\
\textbf{3D Centreline Extraction.} Contour coordinates from vessel segmentation predictions are extracted and the 2D centreline determined. As the geometries of multiple views are known, it is possible to back-project a 2D point (u,v) to the 3D plane (x,y,z) using triangulation \cite{chabi2022self}. 
\\
\textbf{Mesh Estimation.} A mesh is constructed from the 3D centreline vector $P \in \mathbb{R}^{100,3}$ and mean lumen radius vector $R \in \mathbb{R}^{100,1}$ in Blender software. For each 3D centreline point $P_{i}$ and lumen radius $R_{i}$, a ring of 60 vertices $V_{i}$ is created where the distance from $P_{i}$ to each new vertex $V_{i,j}$ is $R_{i}$. The ring is placed on the normal to the 3D centreline, where $\Vec{P_{i}P_{i+1}} \perp \Vec{P_{i}V_{i,j}} $. Each resulting coronary vessel mesh has 6,000 vertices connected by 11,940 edges to form 5,940 quad faces.

\subsection{Surface Refinement}
\label{ssec:surfref}

\textbf{Image Features.} A VGG16 \cite{simonyan2014very} network encodes image features from each view, and returns the outputs from the first four convolutional blocks to the projection layer. This lightweight encoder
is preferred over EfficientNet, as feature map resolution is maintained in the early encoding layers which helps to
preserve the vessel of interest which only occupies a small area of the original image. High level feature maps from late in the network contain more coarse representations of vessel morphology such as curvature, thickness, shape, while low level features from early in the network contain finer details such as edges. The combination of high and low level features ensures both coarse and fine details are leveraged for improved deformation. 
\\
\textbf{Forward-Projection.} Image feature maps and the initialised mesh are passed to a forward-projection layer which maps relevant pixels to vertices on the mesh template. In forward-projection, a 3D point (x,y,z) on the initial mesh is projected to the 2D image feature plane (u,v) for each view. Four adjacent pixels are sampled using bi-linear interpolation for every 2D coordinate on each of the four feature maps, and the output vectors concatenated between views. The 3D positional encoding values for the initial mesh are also concatenated with the output vector resulting in a feature vector of size 483 for each vertex. 
\\
\textbf{Graph Convolutional Network.} Our graph network receives graph features as input from the forward-projection layer where 3D shape information for each vertex of the initial mesh is encoded. Features are passed through a block of 8 GCN layers \cite{kipf2016semi}, where information is aggregated from neighbouring nodes, eventually regressing the 3D location for each vertex. Residual connections are added to minimise the over-smoothing effect common in graph networks. The graph convolutional operation is formalised by: 

\begin{equation} \label{gcn_layer} 
H^{(l+1)} = \sigma (\hat{D}^{-1/2} \hat{A} \hat{D}^{-1/2}H^{(l)}W^{(l)})
\end{equation}

\noindent Where $\hat{A}$ is the adjacency matrix with self-loops of the graph $G$, $\hat{D}_{ii} = \sum_{j}^{} \hat{A}_{ij}$ is the diagonal degree matrix of $G$, $W^{(l)}$ are trainable weights for each layer and $\sigma$ is a ReLU activation. $H^{(l)} \in \mathbb{R}^{N \times D}$  is output for layer $l$ where $N$ is the number of nodes in the graph and $D$ is the dimension of the hidden layer given to be 128 \cite{kipf2016semi}. 
\\
\noindent \textbf{Differentiable Rendering.}
We implement a differentiable rendering pipeline based on Soft-Rasterizer \cite{liu2019soft} which provides an additional 2D supervisory signal during training. At each iteration our 3D mesh prediction is projected to the angiographic image space using the camera geometry of each of two views. Alignment between projection silhouette and vessel mask is calculated using the same $\mathcal{L}_{seg}$ loss from the MI module. 
\\
\noindent \textbf{Losses.} The SR module is supervised by a point-wise mean squared error loss $\mathcal{L}_{MSE}$ to minimise the distance between predicted $\widehat{v}$ and ground truth $v$ mesh vertices.
\begin{equation} \label{mse}
\mathcal{L}_{MSE} = \frac{1}{\left | v \right |} \sum_{i=1}^{{\left | v \right |}} (v_{i} - \widehat{v_{i}})^{2}    
\end{equation}
However, a MSE loss alone is not sufficient to produce realistic vessel shapes with smooth and even surfaces, hence a set of 3 regularization losses must be used to avoid undesirable morphological artefacts. A normal loss $\mathcal{L}_{Norm}$ ensures faces are pointing in the correct direction by penalising large angles between the predicted vertex normals $n_{\widehat{v}_{i}}$ and ground truth vertex normals $n_{v_{i}}$.
\begin{equation} \label{normalloss}
\mathcal{L}_{Norm} = 1 - \sum_{i}^{}  \left \| n_{\widehat{v}_{i}} \cdot n_{v_{i}} \right \|
\end{equation}
\begin{equation}
n_{v_{i}} = \frac{1}{N(f)} \sum_{j=1}^{N(f)}  f_{i, j} 
\end{equation}Where $n_{v_{i}}$ is the normal at a vertex $v_{i}$, $N(f)$ is the number of faces adjacent to the vertex $v_{i}$, $f_{i, j}$ refers to the normal of the $j$th face surrounding the $i$th vertex. The face normal is simply calculated as the cross product between any two vectors made from vertices that make up a face.
\\
\indent An edge loss $\mathcal{L}_{Edge}$ is used to restrict the length of the edge from its initial state and leads to a visually appealing surface, 
\begin{equation} \label{edgeloss}
\mathcal{L}_{Edge} = \frac{1}{\left | E \right |}\sum_{v_{i}, v_{j} \in E}^{}\left \| v_{i} - v_{j} \right \|_{2}^{2}
\end{equation}
\noindent where $E$ is a set of graph edges. 
\\
A Laplacian loss $\mathcal{L}_{Lap}$ is used to penalise overlapping vertices, ensuring a smooth mesh surface. The discrete Laplacian $L$  at vertex $v_{i}$ and the Laplacian loss $\mathcal{L}_{Lap}$ are calculated as follows:
\begin{equation} \label{lapvertex}
L({v_{i}}) = \frac{1}{N(v_{i})} \sum_{j\in N(v_{i})}^{}(v_{i} - v_{j})
\end{equation}
\begin{equation} \label{laploss}
\mathcal{L}_{Lap} = \frac{1}{n} \sum_{i=1}^{n}  \left \| L(v_{i}) - L(\widehat{v_{i}}) \right \| _{2}^{2}
\end{equation}Where $N(v_{i})$ is the number of neighbouring vertices $v_{j}$ to the vertex $v_{i}$ that are connected by an edge, and $L(v_{i})$ refers to the discrete Laplacian at a given vertex $v_{i}$. The losses are combined using Eq. \ref{totalloss}, where weights are found through extensive experimentation. 
\begin{equation} \label{totalloss}
\mathcal{L} = \mathcal{L}_{MSE} + 0.01\mathcal{L}_{Norm} + 2.5\mathcal{L}_{Edge} + 100\mathcal{L}_{Lap} + 0.0002\mathcal{L}_{Seg}
\end{equation}

\begin{figure}[htb]
\begin{minipage}[b]{1.0\linewidth}
  \centering
  \centerline{\includegraphics[width=8.5cm]{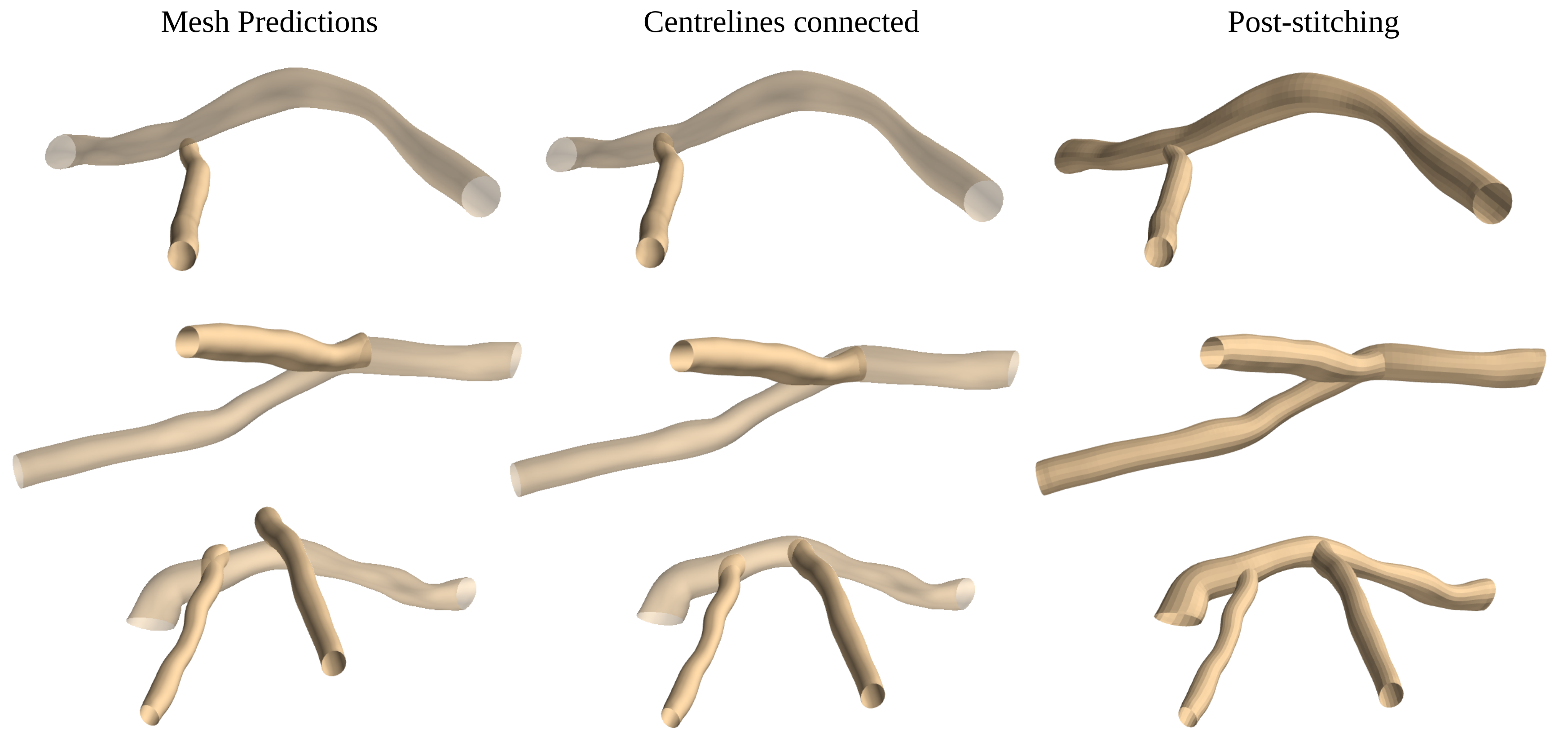}}
\end{minipage}
\caption{Stitching Process.}
\label{stitching}
\end{figure}

\subsection{Stitching}
\label{ssec:stitch}

Main and side branches are reconstructed separately using the above method and then unnormalised to give their original positions in 3D space. It is assumed that the start point of side branch centreline should intersect somewhere along the main branch centreline. However, due to errors in acquisition or reconstruction, there is often a small distance between these. To fix this, side branch(es) are translated in 3D space so that their start point shares the same coordinates with the nearest main branch centreline point. The main and side branch(es) are then combined using a Boolean Union operation which removes any area which is shared between the two mesh, and creates a new mesh from the unshared areas. This is formalised by $A \cup B := \left \{ x \in \mathbb{R}^{3} \mid x\in A \: \textup{and}\:  x\in B\right \}$ where $A$ and $B$ are the main branch and side branch meshes. The full stitching process is illustrated in Fig \ref{stitching}.

\section{Experiments and results}
\label{sec:experiments}

\subsection{Dataset}
\label{ssec:dataset}

\textbf{Overview} A dataset of angiographic studies of 414 patients with atherosclerotic lesions was obtained using four different XCA machines. Coronary reconstruction was performed on a segment of intermediate significance detected on coronary angiography (90\%$>$ Diameter Stenosis $>$30\%). In total 489 segments in these angiographic images fulfilled the inclusion criteria and were included in the final analysis. The dataset was split into 70\% train, 15\% val and 15\% test examples, where each example has a pair of angiographic images and acquisition geometry information. 
\\
\textbf{Ground Truth.} 2D segmentation masks and 3D ground truth meshes were created from two precisely calibrated views using commercially validated QAngio XA 3D RE software (Medis Medical Imaging Systems, Leiden, The Netherlands), with manual correction performed by two expert cardiologists. For bifurcated vessels, only 2D ground truth masks are available. Each mesh is standardised to 6000 vertices, 5940 faces and 11940 edges. 
\\
\textbf{Pre-Processing.} Grayscale angiographic images (512 x 512) from each view were normalised to the pixel values [0,1]. 3D initialisation and ground truth mesh vertices were normalised by a translation to the origin, rotated in 3D to fit the z-axis and normalised to the interval [0,1]. Camera extrinsic and intrinsic parameters found in the acquisition DICOM file were used to roughly calibrate views. Calibration was then optimised to match the vessel start and end point correspondence given by clinicians using a Levenberg–Marquardt algorithm \cite{more1978levenberg}. 

\subsection{Training}
\label{ssec:training}

The MI module was trained using an Adam optimizer with an initial learning rate of 0.001 with an epoch decay factor of 0.96. A batch size of 8 was used and the best validation loss was saved during training over 100 epochs. The SR module was trained using the same set up for 500 epochs, a decay factor of 0.99 and batch size of 1. 3DAngioNet is implemented with PyTorch using an NVIDIA 1080Ti GPU. 

\begin{figure}[htb]
\begin{minipage}[b]{1\linewidth}
  \centering
  \centerline{\includegraphics[width=8.5cm]{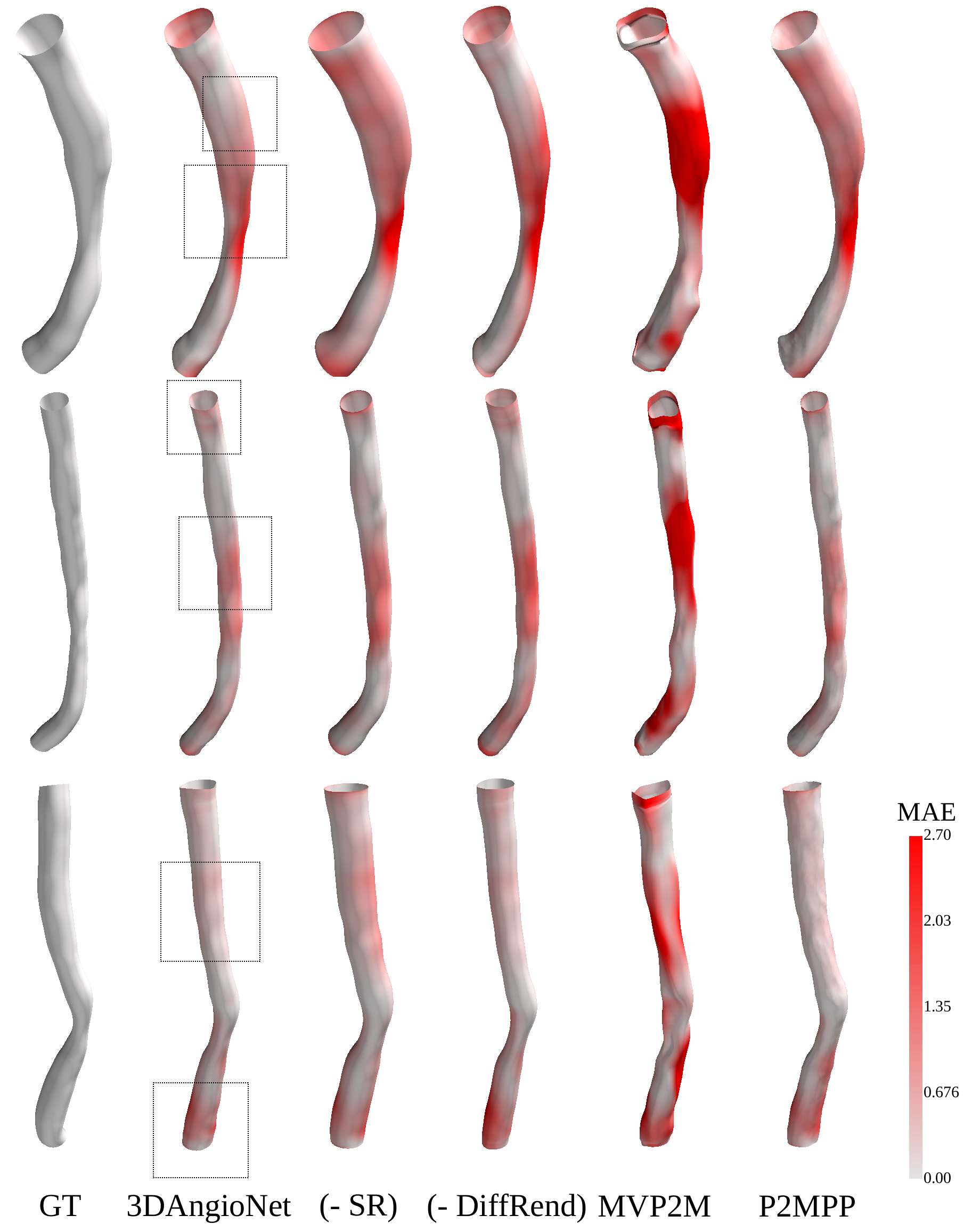}}
\end{minipage}
\caption{3D Qualitative Comparison. Bounding boxes indicate regions of best performance.}
\label{qual_3d}
\end{figure}

\begin{figure}[htb]
\begin{minipage}[b]{1\linewidth}
  \centering
  \centerline{\includegraphics[width=7.5cm]{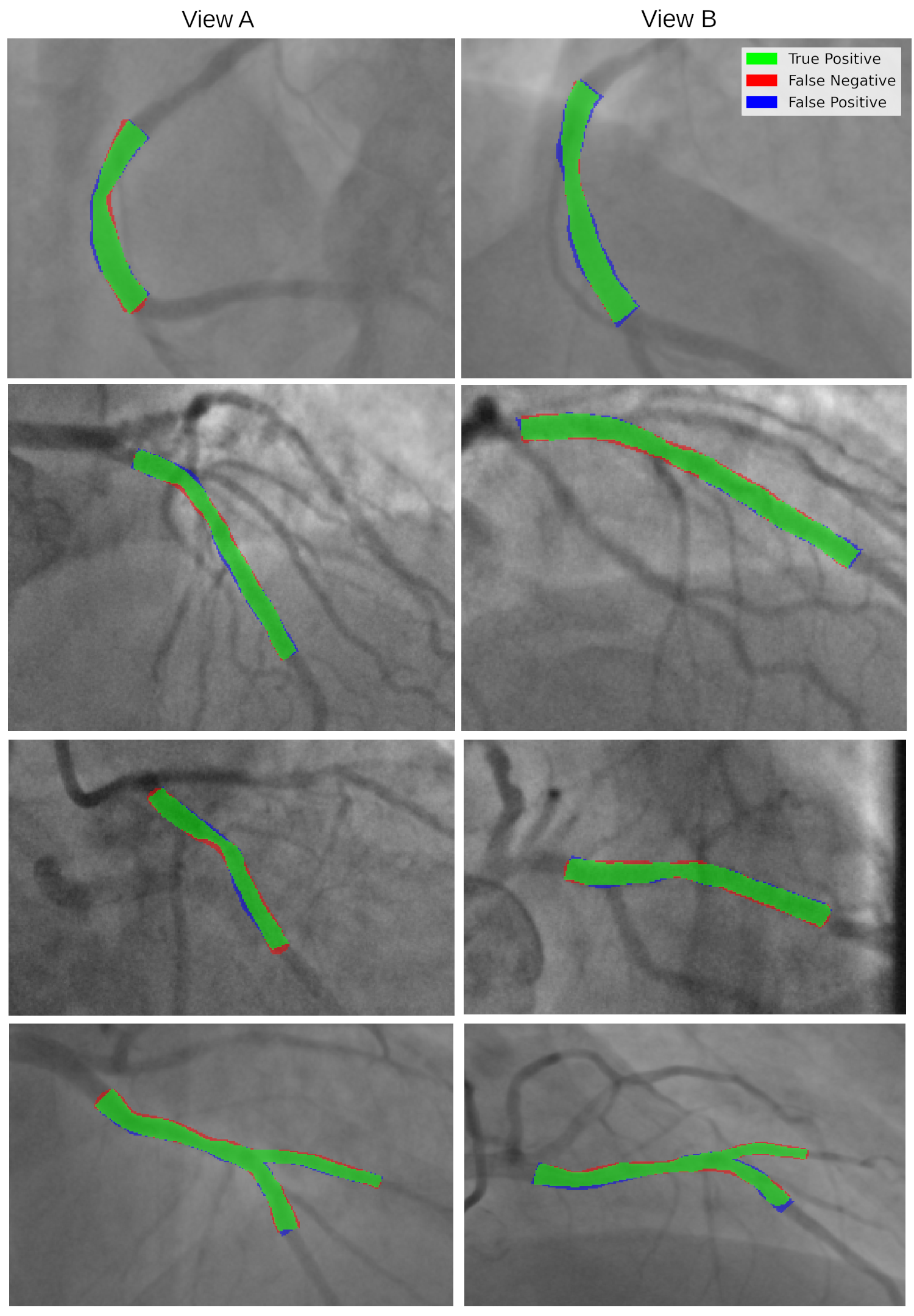}}
\end{minipage}
\caption{2D Qualitative comparison between 2D projections from Fig \ref{qual_3d} and the 2D ground truth (2-3x Zoom). Forth row compares a bifurcated vessel prediction after stitching. This comparison does not appear in Fig.\ref{qual_3d} as only 2D ground truth is available for bifurcated vessels. }
\label{qual_2d}
\end{figure}

\subsection{Comparison to Current Methods}
\label{ssec:comparison}
We validate our framework by comparing test set predictions to the 3D ground truth using mean absolute error (MAE) and Hausdorff distance (HD).  Precision and recall are also calculated by determining the percentage of points that can find a nearest neighbour within a threshold value $\tau$ \cite{wang2018pixel2mesh}. F-score is the harmonic mean between precision and recall, and $\tau$ set at 0.0005. In addition, we project the 3D mesh predictions back to the 2D angiographic image space and calculate alignment with the 2D ground truth using Dice and Jaccard Index.
\\
\indent Results in Fig. \ref{qual_3d}, Fig. \ref{qual_2d} and Table \ref{quant} demonstrate that our model is able to reconstruct vessels with minimal error and high resemblance to those created using state-of-the-art Medis software. By automating the vessel reconstruction step, our framework offers improvements in efficiency with reconstructions taking approximately 5s (1080Ti GPU, Batch size=1), compared to an estimated 4 minutes for Medis. 
\begin{table}[!t]
\centering
\footnotesize
\begin{tabular}{llllll}
\hline
\rule{0pt}{2ex}
                       & MAE$\downarrow$ & HD$\downarrow$ & F-score$\uparrow$ & Dice$\uparrow$ & Jaccard$\uparrow$ \\ 
                       & (mm) & (mm) & & \\\hline
\rule{0pt}{2ex}3DAngioNet      & \textbf{0.3459} & \textbf{1.1884} & \textbf{82.60} & \textbf{87.59} & \textbf{78.57} \\
(-) SR                 & 0.3658 & 1.8536 & 76.64 & 86.62 & 77.01 \\
(-) DiffRend         & 0.3570 & 1.6283 & 81.63 & 86.57 & 76.95 \\ 
\hline
\rule{0pt}{2ex}MVP2M & 0.4266 & 1.7612 & 81.54 & 86.15 & 76.25         \\
P2MPP & 0.3545 & 1.6520 & 81.90 & 87.40 & 78.18 \\
\hline
\rule{0pt}{0.5ex}  
\end{tabular}
\caption{3D and 2D Quantitative Comparison.}
\label{quant}
\end{table}
\\
\indent Further comparison between our system and other existing methodologies is difficult as our dataset is private, and no other public datasets or source code for similar vessel reconstruction methods is available. We compare to general purpose multi-view 3D reconstruction methods Multi-View-Pixel2Mesh (MVP2M) \cite{wen2019pixel2mesh++} and Pixel2Mesh++ (P2MPP) \cite{wen2019pixel2mesh++} networks. As these networks initialise a mesh from an ellipsoid they are incapable of creating hollow tubular structures, therefore we compare performance while using the mesh initialisation from 3DAngioNet. Table \ref{quant} and Fig. \ref{qual_3d} demonstrates that our approach has more accurate and consistent reconstructions with fewer errors in lumen surface.
\\
\indent Table \ref{quant} presents an ablation study which compares performance using a number of different strategies: (1) Removing the SR module, and creating a mesh using the MI module only, and (2) Removing the differentiable rendering unit and additional 2D supervision. Models trained using differentiable rendering and SR module outperform those without, justifying our system design choices. 
\section{Conclusion}
\label{sec:typestyle}
In this work we propose an effective method for 3D coronary vessel reconstruction. We exploit segmentation and graph learning to automate this process, with minimal error. Experimental results demonstrate that our method is superior to general purpose 3D reconstruction algorithms, while producing mesh vessels with high resemblance to those created manually using state-of-the-art software. This work enables streamlining of time-consuming tasks while opening up future research avenues in atherosclerotic plaque progression. Our model is validated on data from four different XCA machines, demonstrating that our approach is generalizable to new data. In future work we aim to fully automate 3DAngioNet by learning the SOI region rather than using clinical guidance, and improve lumen resolution by co-registering QCA with intravascular imaging modalities. 

\section{Compliance with ethical standards}
\label{sec:ethics}
This research study was conducted retrospectively using human subject data made available by Barts Health NHS Trust. Ethical approval was not required as confirmed by the license attached with the data.

\section{Acknowledgments}
\label{sec:acknowledgments}

This research is part of AI-based Cardiac Image Computing (AICIC) funded by the faculty of Science and Engineering at Queen Mary University of London. We would like to thank Medis Medical Imaging Systems for access to their QAngio XA
3D RE software which was used to create ground truth standards. 

\bibliographystyle{IEEEbib}
\bibliography{strings,refs}

\end{document}